\documentclass[12pt]{article}
\title{\bf Comment on the proper QCD string dynamics in a heavy-light system}
\author{A. V. Nefediev\thanks{e-mail: nefediev@heron.itep.ru}}
\date{\small\it Institute of Theoretical and Experimental Physics,\\
B.Cheremushkinskaya 25, 117218, Moscow, Russia}
\newcommand{\be}{\begin{equation}}
\newcommand{\ee}{\end{equation}}
\newcommand{\vx}{\vec{x}}
\newcommand{\vy}{\vec{y}}

\newcommand{\vp}{\vec{p}}

\newcommand{\ds}{\displaystyle}
\begin{document}
\maketitle

\begin{abstract}
The string correction to the inter-quark interaction at large distances is derived
using the field theory approach to a heavy-light quark-antiquark system in the modified Fock-Schwinger
gauge. 
\end{abstract}

Quantum chromodynamics at large distances is believed to be a string theory with the effective extended object
--- the QCD string --- formed by nonperturbative gluons, which plays an important role in hadronic phenomenology.
It was demonstrated in a number of approaches that the account for the proper dynamics of the QCD string
strongly affects hadronic spectra and is necessary to explain the correct Regge 
trajectory slopes \cite{Olsson,DKS,regge}, to resolve puzzles with the identification of new states \cite{mesons}, 
and so on. At large inter-quark distances this dynamics can be encoded in the so-called 
string correction \cite{MP,DKS}, well known in the theory of the straight-line Nambu-Goto string with the tension
$\sigma$ and with massive quarks at the ends. The Lagrangian of this system is
\be
L=-M\sqrt{\dot{x}_1^2}-m\sqrt{\dot{x}_2^2}-\sigma\int_0^1d\beta\sqrt{(\dot{w}w')^2-\dot{w}^2w'^2},\quad
w_{\mu}(t,\beta)=\beta x_{1\mu}(t)+(1-\beta)x_{2\mu}(t),
\label{NG}
\ee
which leads to the c.m. Hamiltonian, at large inter-quark distances $r=|\vec{x}_1-\vec{x}_2|$, for $M\gg m$, \cite{DKS}:
\be
H\approx M+m+\frac{\vec{p}^2}{2m}+\sigma r-\frac{\sigma \hat{L}^2}{6m^2r}+\ldots,
\label{H}
\ee
where, for the sake of convenience, we synchronised the quark times, $x_{10}=x_{20}\equiv x_0$, and 
fixed the reparametrisation invariance of the Lagrangian (\ref{NG}) by the laboratory frame condition 
$t=x_0$. The nonperturbative spin-orbit interaction comes from the area law for the Wilson loop \cite{3},
\be
V_{so}=-\frac{\vec{\sigma}\vec{L}}{4m^2r},
\label{dH}
\ee
and should be added to the Hamiltonian (\ref{H}). The expansion in Eq.~(\ref{H}) is valid for $m\gg\sqrt{\sigma}$. 
Perturbative Coulomb interaction as well as extra spin-dependent terms due to the latter 
can be taken into account, and the
resulting model appears rather successful in describing hadronic spectra (see, for example, \cite{spec1},
where the Hamiltonian (\ref{H}) supplied by the perturbative interaction, but without the string correction, 
was used). A more sophisticated approach based
on the einbein field formalism \cite{einbein} is also well known in the literature \cite{DKS}. This method 
possesses several advantages as compared to the Hamiltonian (\ref{H}) since, in this case, the corresponding 
Hamiltonian is given in terms of the effective dynamically generated quark masses $\mu$'s, given by the 
extremal values of the corresponding einbeins \cite{Green}. For light quarks such a dynamical mass 
appear of order of the interaction scale, $\mu\sim\sqrt{\sigma}$, that is, much larger than the current 
quark mass --- the latter can be even put to zero.  

Recently, another approach to heavy-light systems was
suggested, based on the Schwin\-ger-Dyson series for a light quark in presence of a static antiquark
\cite{SN,SN2}. Namely, the Schwinger-Dyson equation, in Euclidean space-time,
\be
(-i\hat{\partial}_x-im)S(x,y)-i\int d^4zM(x,z)S(z,y)=\delta^{(4)}(x-y),
\label{11}
\ee
was derived in the modified Fock--Schwinger gauge \cite{10}, 
\be
A_4(x_4,\vec{0})=0,\quad\vec{x}\vec{A}(x_4,\vec{x})=0,
\label{5}
\ee
where the self--energy part $M(x,z)$ and the light-quark Green's function (also playing the role of the
$q\bar{Q}$ Green's function) are given by \cite{SN}
\be
-iM(x,z)=K_{\mu\nu}(x,z)\gamma_{\mu}S(x,z)\gamma_{\nu},\quad
S(x,y)=\frac{1}{N_C}\langle\psi^{\beta}(x)\psi^+_{\beta}(y)\rangle.
\label{12}
\ee
 
The interaction kernel $K_{\mu\nu}$ can be expressed in terms of the irreducible field strength 
correlator $\langle F^a_{\mu\nu}(x)F^b_{\lambda\rho}(y)\rangle$ \cite{4},
\be
\langle F^a_{\mu\nu}(x)F^b_{\lambda\rho}(y)\rangle=\delta^{ab}\frac{2N_C}{N_C^2-1}D(x_0-y_0,|\vx-\vy|)
(\delta_{\mu\lambda}\delta_{\nu\rho}-\delta_{\mu\rho}\delta_{\nu\lambda})+
\Delta^{(1)},
\label{15}
\ee
where the second term $\Delta^{(1)}$ is a full derivative and does not contribute to confinement. 
As we are interested in the long-range force, we consider only the term proportional to $D(x-y)$ in (\ref{15}) 
which, in contrast to $\Delta^{(1)}$, contributes to the area law with the string tension 
\be
\sigma=2\int_{0}^\infty d\tau\int_{0}^\infty d\lambda D(\tau,\lambda).
\label{16}
\ee

Finally, for the kernel $K_{\mu\nu}$ in the gauge (\ref{5}), one has $(\tau=x_4-y_4)$ \cite{SN,SN2}:
\be
\begin{array}{l}
K_{44}(\tau,\vx,\vy)=(\vx\vy)\ds\int_0^1d\alpha\int_0^1 d\beta 
D(\tau,|\alpha\vx-\beta\vy|),\\
{}\\
K_{i4}(\tau,\vx,\vy)=K_{4i}(\tau,\vx,\vy)=0,\\
{}\\
K_{ik}(\tau,\vx,\vy)=((\vx\vy)\delta_{ik}-y_ix_k)\ds
\int_0^1\alpha d\alpha\int_0^1 \beta d\beta 
D(\tau,|\alpha\vx-\beta\vy|).
\end{array}
\label{Ks}
\ee

Using a consequent expansion of Eq.~(\ref{11}) for a large quark mass $m$ ($m\gg\sqrt{\sigma}$ and 
$mT_g\gg1$ \cite{SN2,we1}, where $T_g$ is the gluonic correlation length) one can derive the inter-quark 
interaction which is in agreement with the Eichten--Feinberg--Gromes results \cite{1,2}. Then,
applying the Foldy--Wounthuysen (FW) transformation to the resulting
interaction, it is easy to derive a Hamiltonian of the heavy-light system, at $r\gg T_g$, in the form 
\cite{SN2,we1}:
\be
H_{FW}=M+m+\frac{\vp^2}{2m}+\sigma r-\frac{\vec{\sigma}\vec{L}}{4m^2r}+\ldots, 
\label{H2}
\ee
where the ellipsis denotes terms $O\left(\frac{\sigma r}{mT_g}\right)$ suppressed in the limit 
$mT_g\gg 1$ \cite{we1,magn}. 

The Hamiltonian (\ref{H2}) coincides with the Hamiltonian of the quantum-mechanical
quark-antiquark system connected by the Nambu-Goto string supplied by the nonperturbative spin-dependent
interaction given by Eqs.~(\ref{H}), (\ref{dH}). In the meantime, an important ingredient mentioned
above --- the 
string correction --- is still missing in the formula (\ref{H2}). The aim of the present paper is to resolve this
inconsistency and, therefore, to complete matching of the two approaches: one, based on the quantum mechanical
string model, and the other, following from the field theoretical treatment of the heavy-light quark-antiquark system.

Following the path integral ideology, we consider the trajectory of the quark, $\vec{r}(t)$,
such that the two consequent positions of the latter are $\vec{x}=\vec{r}(t_1)$ and $\vec{y}=\vec{r}(t_2)$.
Therefore for close $t_1$ and $t_2$ one can use the expansion:
\be
\vec{y}=\vec{r}(t_2)=\vec{r}(t_1+\tau)\approx\vec{r}(t_1)+\dot{\vec{r}}(t_1)\tau=
\vec{x}+\frac{\vec{p}}{m}\tau,
\label{exp}
\ee
where $\vec{p}$ is the momentum of the quark. Due to the rotational invariance the function $D$ from
Eq.~(\ref{15}) actually depends on the certain combination of its arguments, 
$D(\tau,\lambda)=D(\tau^2+\lambda^2)$. In our case $\tau=t_2-t_1$ and $\lambda=|\alpha\vec{x}-\beta\vec{y}|$,
so that, with the help of the expansion (\ref{exp}), one easily finds:
\be
\tau^2+\lambda^2=\tau^2+\left[(\alpha-\beta)\vec{r}+\alpha\tau\frac{\vec{p}}{m}\right]^2
=\left[1+\frac{\alpha^2p^2}{m^2}\right](\tau-\tau_0)^2+\frac{(\alpha-\beta)^2}
{1+\frac{\alpha^2p^2}{m^2}}
\left[r^2+\alpha^2\frac{L^2}{m^2}\right],
\ee
where $\vec{r}\equiv\vec{x}$, $\vec{L}$ is the angular momentum, $\vec{L}=[\vec{r}\times\vec{p}]$, and the constant
$\tau_0=\frac{\alpha(\beta-\alpha)(\vec{r}\vec{p})}{m(1+\alpha^2p^2/m^2)}$ can be 
excluded using an appropriate shift of the time variable $\tau$, so we omit it below. 

The confining spin-independent interaction, at large inter-quark distances and in the limit $mT_g\gg 1$, 
is given by the formula \cite{SN,SN2,we1}:
\be
V_{conf}(r)=\gamma_{\mu}\frac{1+\gamma_0}{2}\gamma_{\nu}\int_0^\infty d\tau 
K_{\mu\nu}(\tau,\vx,\vy)_{\left|\vy\to\vx\right.}.
\label{Vc}
\ee
Using the relations (\ref{Ks}), one can easily calculate that  
$$
\int_0^\infty d\tau K_{00}(\tau,\vx,\vy)_{\left|\vy\to\vx\right.}
=r^2\int_0^{\infty}d\tau\int_0^1d\alpha\int_0^1 d\beta 
D\left(\tau\sqrt{1+\frac{\alpha^2p^2}{m^2}},(\alpha-\beta)\sqrt{
\frac{r^2+\frac{\alpha^2L^2}{m^2}}{1+\frac{\alpha^2p^2}{m^2}}}\right)
$$
\be
\mathop{\approx}\limits_{r\gg T_g}
r\left(2\int_0^\infty d\tau'\int_0^\infty
d\lambda D(\tau',\lambda)\right)\int_0^1\frac{d\alpha}{\sqrt{1+\frac{\alpha^2L^2}{m^2r^2}}}
\approx\sigma r-\frac{\sigma L^2}{6m^2r},
\ee
and, similarly,
\be
\int_0^\infty d\tau K_{ik}(\tau,\vx,\vy)_{\left|\vy\to\vx\right.}
\approx(\delta_{ik}-n_in_k)\left(\frac13\sigma r-\frac{\sigma L^2}{10m^2r}\right),
\ee
where the definition of the string tension (\ref{16}) was used, as well as the case $n=0$ of 
the general formula
$$
\int_0^1d\alpha\int_0^1d\beta f(\alpha,\beta)(\alpha-\beta)^nD(\tau,|\alpha-\beta|a)
\mathop{\approx}\limits_{a\gg 1}\frac{2}{a^{n+1}}\int_0^{\infty}d\lambda
D(\tau,\lambda)\int_0^1d\alpha f(\alpha,\alpha),
$$
which holds for an arbitrary function $f(\alpha,\beta)$, provided $f(\alpha,\alpha)\neq 0$. 

Therefore, the confining interaction (\ref{Vc}), in the limit $mT_g\gg 1$, becomes
\be
V_{conf}(r)=\left(\frac56+\frac16\gamma_0\right)\sigma
r-\left(\frac{11}{60}-\frac{1}{60}\gamma_0\right)\frac{\sigma L^2}{m^2 r},
\label{VV1}
\ee
or, after the Foldy--Wounthuysen rotation this corresponds to the confining potential
\be
V_{conf}^{FW}(r)=\sigma r-\frac{\sigma L^2}{6m^2r}.
\label{VV2}
\ee

The first term of the interaction (\ref{VV1}), (\ref{VV2}) was obtained in refs.~\cite{SN,SN2,we1,magn}, 
whereas the second term was
missing due to the immediate substitution of $\vy=\vx$ in the formula (\ref{Vc}), which holds up to the order
$1/m$, but, as demonstrated above, fails in the next order, $1/m^2$. As a result, the string correction was
lost, although a more accurate expansion of the correlator $D$, performed in this paper, allows one to
reproduce the confining potential, including its part due to the proper string dynamics. Therefore we
conclude that indeed the string correction accompanies the linear confinement potential whatever approach is
used to derive the latter, provided the string picture of confinement is adopted. Meanwhile the suggested
approach is rather inconvenient for further investigations of the confining interaction in the approach of the
Schwinger-Dyson nonlinear equation (\ref{11}). On the other hand, a promising step is made in the paper
\cite{gg} where a contour gauge is introduced which generalises the gauge condition (\ref{5}) for the case of
an arbitrary trajectory of the heavy particle. Formally, Eqs.~(\ref{11}), (\ref{12}) remain valid, though the
kernel of the interaction becomes contour-dependent. In the meantime, the form of the contour depends on the
heavy particle trajectory, that is, it is defined dynamically, and the problem becomes selfconsistent.
Consequent expansion of the aforementioned contour around the straight-line form may provide a way of 
systematic account for the $(1/m)^n$ and $(1/M)^n$ corrections in this approach.
\medskip

Useful discussions with Yu. A. Simonov and Yu. S. Kalashnikova are acknowledged.
This work is supported by INTAS, via grants OPEN 2000-110 and YSF 2002-49, by the grant NS-1774.2003.2, 
and by the Federal Programme of the Russian Ministry of Industry, Science and Technology No 40.052.1.1.1112.

\end{document}